\documentclass[prb,twocolumn, showpacs,preprintnumbers,amsmath,amssymb, superscriptaddress]{revtex4}

\usepackage{graphicx}
\usepackage{dcolumn}
\usepackage{bm}

\begin{document}


\title{
Strain effects on ferroelectric polarization and magnetism in orthorhombic HoMnO$_{3}$ \\
}

\author{Diana Iu\c{s}an}
\email[]{diana.iusan@spin.cnr.it}
\affiliation{Department of Physics and Astronomy, Uppsala University, Box 516, 75120 Uppsala, Sweden}
\affiliation{Consiglio Nazionale delle Ricerche - Superconducting and Innovative Materials and Devices (CNR-SPIN), 67100 L'Aquila, Italy}
\author{Kunihiko Yamauchi}
\affiliation{Consiglio Nazionale delle Ricerche - Superconducting and Innovative Materials and Devices (CNR-SPIN), 67100 L'Aquila, Italy}
\affiliation{Institute of Scientific and Industrial Research (ISIR-SANKEN), Osaka University, 8-1 Mihogaoka, Ibaraki, Osaka, 567-0047, Japan}
\author{Paolo Barone}
\affiliation{Consiglio Nazionale delle Ricerche - Superconducting and Innovative Materials and Devices (CNR-SPIN), 67100 L'Aquila, Italy}

\author{Biplab Sanyal}
\affiliation{Department of Physics and Astronomy, Uppsala University, Box 516, 75120 Uppsala, Sweden}
\author{Olle Eriksson}
\affiliation{Department of Physics and Astronomy, Uppsala University, Box 516, 75120 Uppsala, Sweden}
\author{Gianni Profeta}
\affiliation{Consiglio Nazionale delle Ricerche - Superconducting and Innovative Materials and Devices (CNR-SPIN), 67100 L'Aquila, Italy}

\author{Silvia Picozzi}
\affiliation{Consiglio Nazionale delle Ricerche - Superconducting and Innovative Materials and Devices (CNR-SPIN), 67100 L'Aquila, Italy}%

\date{\today}

\newcommand{\hmo}{HoMn$_{}$O$_{3}$}
\newcommand{\rmo}{$R$Mn$_{}$O$_{3}$}

\begin{abstract}
Aiming at increasing the ferroelectric polarization in AFM-E ortho-\hmo, we investigate the in-plane strain effects on both the magnetic configuration and the polarization by means of density functional theory calculations and model Hamiltonian approaches. 
Our results show that the net polarization is largely enhanced under compressive strain, due to an increase of the electronic contribution to the polarization, whereas the ionic contribution is found to decrease.
We identify the electron-lattice coupling, due to Jahn-Teller (JT) distortions, and its response to strain to be responsible for the observed behavior.
The JT-induced orbital ordering of occupied Mn-e$_g^1$ electrons in alternating $3x^2-r^2/3y^2-r^2$ orbital states at equilibrium changes to a mixture with $x^2-z^2/y^2-z^2$ states under in-plane compressive strain. 
The asymmetric hopping of e$_g$ electrons between Mn ions along zig-zag spin chains (typical of the AFM-E spin configuration) is therefore enhanced under strain, explaining the large value of the polarization. 
We reproduce the change in the orbital ordering pattern in a degenerate double-exchange model supplemented with electron-phonon interaction. 
In this picture, the orbital ordering change is related to a change of the Berry phase of the e$_g$ electrons, which in turn causes an increase of the polarization, whose origin is purely electronic. 
\end{abstract}


\maketitle

\section{\label{sec:intro}Introduction\protect\\}

Orthorhombic rare-earth manganites \rmo, which crystalize in the $Pbnm$ structure, represent a prototypical class of ``improper ferroelectrics'',\cite{impro1,impro2,impro3} where electric dipoles are induced by a frustrated magnetic ordering. 
Among this family of compounds, TbMnO$_{3}$ and DyMnO$_{3}$ show non-collinear magnetic structures, where spin-orbit coupling (SOC) via the Dzyaloshinskii-Moriya (DM) interaction is responsible for the weak ferroelectric polarization ($P < \rm{0.1~\mu C/cm^{2}}$) and the magnetoelectricity.\cite{tbmno3.1,tbmno3.2} 
On the other hand, the onset of a ferroelectric state induced by an exchange-striction mechanism (such effect is often called ``inverse Goodenough-Kanamori interaction''\cite{iGK}) was theoretically proposed for HoMnO$_{3}$ in the antiferromagnetic E-type (AFM-E) spin configuration, \cite{sergienko.prl, silvia.prl, kunihiko.prb, Picozzi-JPCM2009} and later experimentally confirmed.\cite{hmo.exp1,hmo.exp2,hmo.exp3}
In the framework of the degenerate double exchange model,\cite{sergienko.prl} it has been shown that a {\it displacement mechanism}, based on the difference - in terms of bond-angles and bond-lengths - between Mn atoms with parallel and antiparallel spins, takes place for inequivalent bridge oxygens, leading to a net ferroelectric polarization ($P$) directed along the $a$ axis.  
The mechanism was confirmed by a density functional theory (DFT) study,\cite{silvia.prl} which also reported a comparable contribution arising from a purely {\it quantum mechanical mechanism}, resulting in a large value of the total ferroelectric polarization, $\rm {\it P} \approx 6~\mu C/cm^{2}$. 
Further calculations based on the Heyd-Scuseria-Ernzerhof hybrid functional report a value of the ferroelectric polarization of $\rm {\approx 2~\mu C/cm^2}$.\cite{Stroppa} 
Microscopically, the electronic contribution is explained in terms of the asymmetric hopping of Mn-$e_{g}$ electrons between neighboring Mn ions, where the ``one-way" direction of hopping is determined both by the orbital ordering and by the E-type spin configuration.\cite{kunihiko.prb} 
As a second effect, the ions  move to enhance the polarization induced by the asymmetric hopping; in fact, the Mn-O-Mn angle between the ferromagnetically coupled Mn spins increases after ferroelectric ionic relaxations, resulting in an increase of the $e_{g}$ hopping integrals. 

For an easier detection of the polarization in experiments, an increase of the polarization in this system is desired. 
A possible path to achieve this is by increasing the Mn-$e_{g}$ hopping amplitude by modulating the $(ab)$ in-plane Mn-O-Mn bond angle $\phi$ (average value of $\phi$=144$^{\circ}$ for HoMnO$_{3}$\cite{Alonso}), as well as the Mn-O bond lengths. 
However, the enhancement of $\phi$ may lead, according to the Goodenough-Kanamori rule,\cite{GK} to the stabilization of other magnetic phases, $i.e.$ the non-collinear configuration, with a smaller $P$, observed in TbMnO$_{3}$ ($\phi$=145$^{\circ}$)
and the paraelectric AFM-A phase in LaMnO$_{3}$ ($\phi$=155$^{\circ}$).
Another possible recipe to increase the polarization $P$ is to shorten the bond lengths in the AFM-E magnetic ground state, hopefully keeping the angle $\phi$ unchanged in such a way that the AFM-E phase does not turn unstable.    
In this context, here we theoretically investigate strain effects both on the magnetic configuration and on the polarization for orthorhombic \hmo, where the in-plane lattice constants $a$ and $b$ are artificially modified in order to mimic the compressive/tensile modes acting in thin films grown on different substrates. 

\section{\label{sec:method}Structural and computational details\protect\\}
Orthorhombic \hmo crystalizes in the $Pbnm$ structure (corresponding to $Pnma$ in the standard setting; in this paper, we chose $c$ as the longest axis).
Here, strong distortions with respect to the ideal cubic perovskite, $i. e.$ GdFeO$_{3}$ type tilting of MnO$_{6}$ octahedral cages and Jahn-Teller (JT) distortions, stabilize the orbital ordering of Mn-$e_{g}^{1}$ states, where the orbital order is expressed as $(3x^{2}-r^{2})/(3y^{2}-r^{2})$. 
Although the hexagonal non-perovskite phase is more stable than the orthorhombic phase in \hmo, the transition to the metastable orthorhombic phase can be obtained by high-pressure synthesis. 

\begin{figure}[t] \vspace{3mm}
\centerline{\includegraphics[width=0.85\columnwidth]{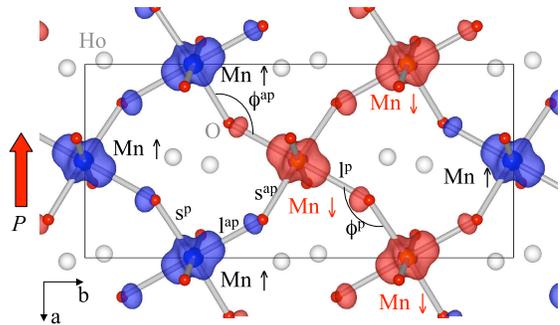}}
\caption{\label{fig.spnchg} 
(Color online) Spin density of the Mn-$e_{g}$ states (within an energy range up to 1 eV below the Fermi energy) in AFM-E configuration without strain, projected in the $(ab)$ plane, referred in the text as {\it in-plane}. 
Blue (dark) and red (light) isosurface show up- and down-spin state (at $\pm$0.1 e$^{-}$), respectively. 
Due to the exchange striction, the Mn-O-Mn bond angles $\phi^{\rm p}$ between parallelly-coupled Mn-spins and  $\phi^{\rm ap}$ between anti-parallelly-coupled Mn-spins are different, as well as the Mn-O bond distances $l^{\rm p}$, $l^{\rm ap}$ and $s^{\rm p}$, $s^{\rm ap}$. The direction of the net polarization ($P \parallel -a$) is denoted by a big red arrow. }
\end{figure}

DFT simulations were performed using the VASP code \cite{vasp} and the PAW pseudopotentials \cite{paw} within the GGA formalism.  
(Note that for ortho-\rmo, it is already known that the GGA+$U$ treatment\cite{ldau} worsens the agreement with experimental results compared to bare GGA, as regards the optimized atomic structure and the magnetic stabilization.\cite{kunihiko.prb, silvia.rmno3.prb})  
The cut-off energy for the plane-wave expansion  of the wave-functions was set to 400 eV and a {\bf k}-point shell of (4, 2, 3) was used for the Brillouin zone integration according to the Monkhorst-Pack special point scheme. 
The in-plane lattice constants corresponding to the unstrained structure are taken as $a$~=~5.2572~\AA\ and $b$~=~5.8354~\AA, equal to the experimental values.\cite{Alonso} 
In order to mimic strain, these values were progressively changed from -4\% (compressive strain) to 4\% (tensile strain). 
The lattice constant $c$ was optimized for each (a,b) value, as well as the internal coordinates. In the unstrained case, the theoretically optimized $c$ is found to be equal to 7.3952~\AA, slightly underestimated compared to the experimental value of 7.3606~\AA. 
In the AFM-E magnetic configuration (AFM $q$ vector along $b$), the unit cell is doubled along $b$ (40 atoms/cell). Therefore, the crystal structure is optimized in the non-centrosymmetric $P2_{1}mn$ symmetry, allowing for a ferroelectric polarization along x ($P_{x}$). 
(Without spin-orbit coupling, the magnetic group is not considered.)
A schematic representation of the structure in the AFM-E magnetic configuration is shown in Fig. \ref{fig.spnchg}.
The electronic polarization $P_{\rm Berry}$=$P_{\rm tot}$ is calculated using the Berry phase method\cite{berry} and compared to the classical point charge model (PCM) result, where $P_{\rm PCM}$=$P_{\rm ion}$ is calculated as the product of the nominal ionic valence and of the ionic displacements.
In order to discuss the stability of different magnetic configurations, the internal atomic coordinates were fully optimized in the AFM-A configuration, characterized by in-plane ferromagnetic and inter-plane antiferromagnetic couplings. 
Using this relaxed centrosymmetric structure, the total energies of several magnetic configurations (antiferromagnetic AFM-A, AFM-C, AFM-G, AFM-E, AFM-E$^{*}$ and ferromagnetic FM configurations)
were compared, and the exchange coupling integrals $J_{ij}$ were calculated by mapping the system onto a Heisenberg model (see Ref. \onlinecite{kunihiko.prb} for details). 
%

\section{\label{sec:struc}Strain effect on atomic structure\protect\\}
\begin{figure}[b] \vspace{3mm}
\centerline{\includegraphics[width=0.84\columnwidth]{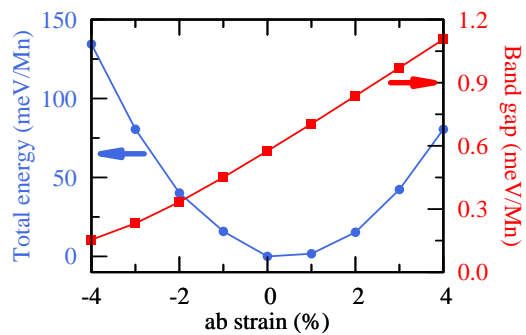}}
\caption{\label{fig.en.OptE} 
(Color online) Calculated variation of the total energy difference and the band gap with optimized atomic structure in AFM-E configuration under in-plane strain. 
 }
\end{figure}

Fig. \ref{fig.en.OptE} shows the dependence of the total energy on strain. 
The deviation from the origin of the energy minimum, giving an estimate of the error of bare GGA treatment with respect to experimental results, is less than 1\% (rather typical for DFT calculations). 
The band gap, given by the JT-splitting of Mn-$e_{g}$ states, is also found to decrease under compressive strain in the $(ab)$ plane.
\begin{figure}[thb] \vspace{3mm}
\centerline{\includegraphics[width=0.84\columnwidth]{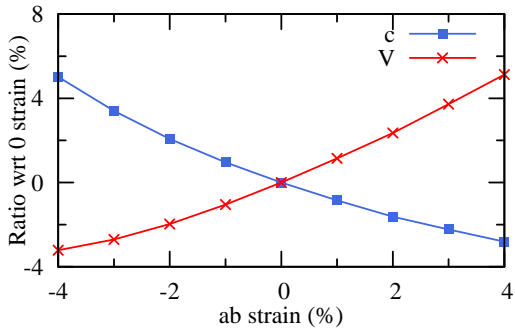}}
\caption{\label{fig.c} 
(Color online) Calculated variation of equilibrium $c$ and lattice volume $V$ under in-plane strain, $a$ and $b$ being fixed for a given strain. 
 }
\end{figure}
\begin{figure}[h] \vspace{3mm}
\centerline{\includegraphics[width=0.84\columnwidth]{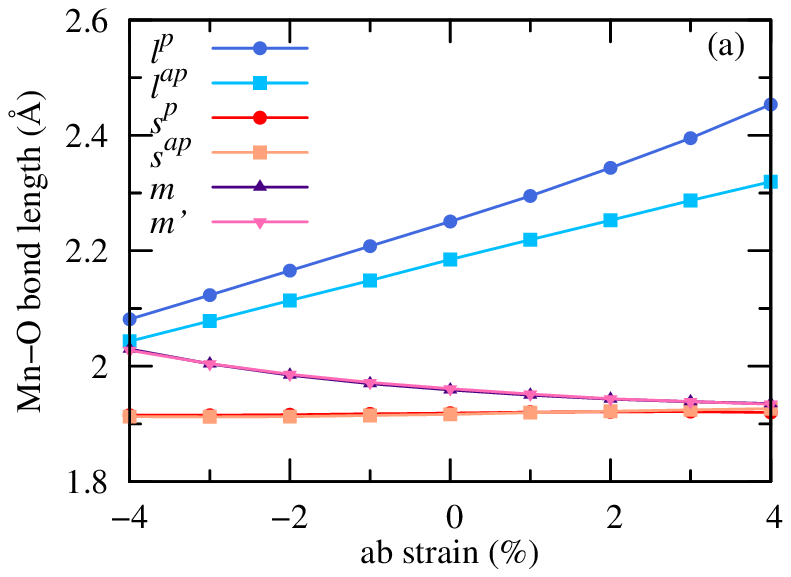}}
\centerline{\includegraphics[width=0.84\columnwidth]{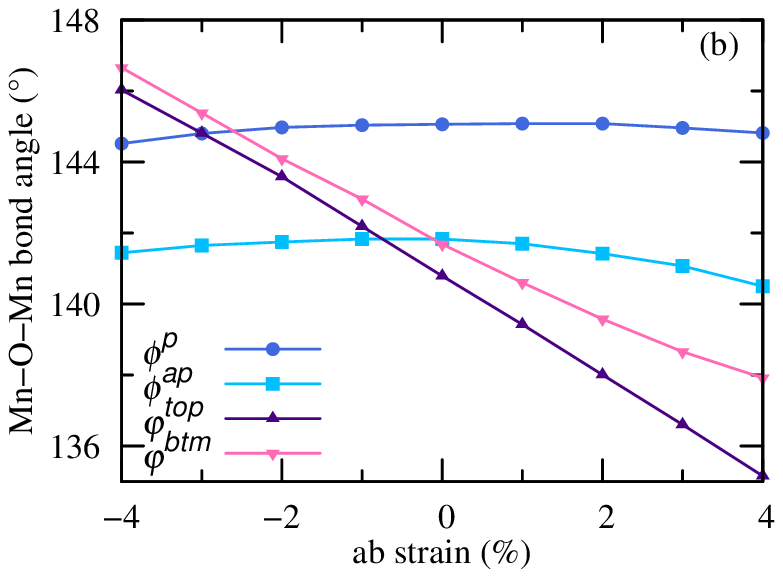}}
\caption{\label{fig.lmsang} 
(Color online) Variation of (a) the Mn-O bond lengths, in-plane $l$, $s$ (long axis and short axis in JT-distorted MnO$_{6}$ octahedron) and inter-plane $m$ and (b) the Mn-O-Mn in-plane bond angles $\phi$ and inter-plane bond angles $\varphi$ in the AFM-E configuration under $(ab)$ in-plane strain.
 }
\end{figure}

In Fig. \ref{fig.c} and Fig. \ref{fig.lmsang} we show how the change in the  $a$ and $b$ lattice constants affects the optimized value of the lattice constant $c$ and the lattice volume, as well as  the Mn-O bond lengths and Mn-O-Mn bond angles in the AFM-E configuration. 
Under in-plane compressive strain, the $c$ lattice constant is expanded, while the volume is contracted. 
The Mn-O bond lengths and the Mn-O-Mn bond angles depend on the relative alignment of the Mn spins, parallel or antiparallel, as a consequence of exchange striction. 
Remarkably, the in-plane bond angles, $\phi^p$ and $\phi^{ap}$, are kept rather constant under strain, whereas the bond lengths are much more modulated. 
The inter-plane Mn-O-Mn bond angles, $\phi^{top}$ and $\phi^{btm}$, show a rapid increase under compressive strain. 
Recalling the fact that the hopping integral between Mn-$e_{g}$ states depends both on the bond angle $\phi$ and on the bond length $d$ --- the overlap integral $\tilde b$ is described as\cite{zhou.goodenough.prl} 
$\tilde b=\cos ((180^{\circ}-\phi)/2)/d^{3.5}$ ---, apparently oxygen ions are displaced in such a way that the hopping energy is maximized. 
On the other hand, the difference  $|\phi^{\rm p}$-$\phi^{\rm ap}|$ of the bond-angle between Mn ions with parallel and antiparallel spins (proven to be relevant for the ionic contribution to the polarization) slightly increases under tensile strain, probably because of the smaller elastic energy that binds oxygen atoms in a relatively larger space in the lattice. 
Such a reduction of the elastic energy is expected to enhance ionic displacement and, consequently, the ionic contribution to polarization, $P_{\rm ion}$.

\section{\label{sec:magn}Magnetic configuration\protect\\}

\begin{figure}[p] \vspace{3mm} 
\centerline{\includegraphics[width=0.85\columnwidth]{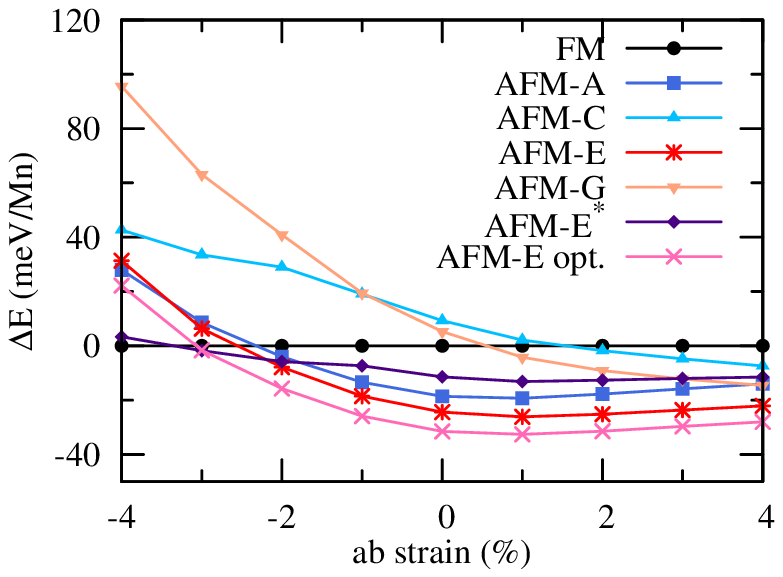}}
\centerline{\includegraphics[width=0.85\columnwidth]{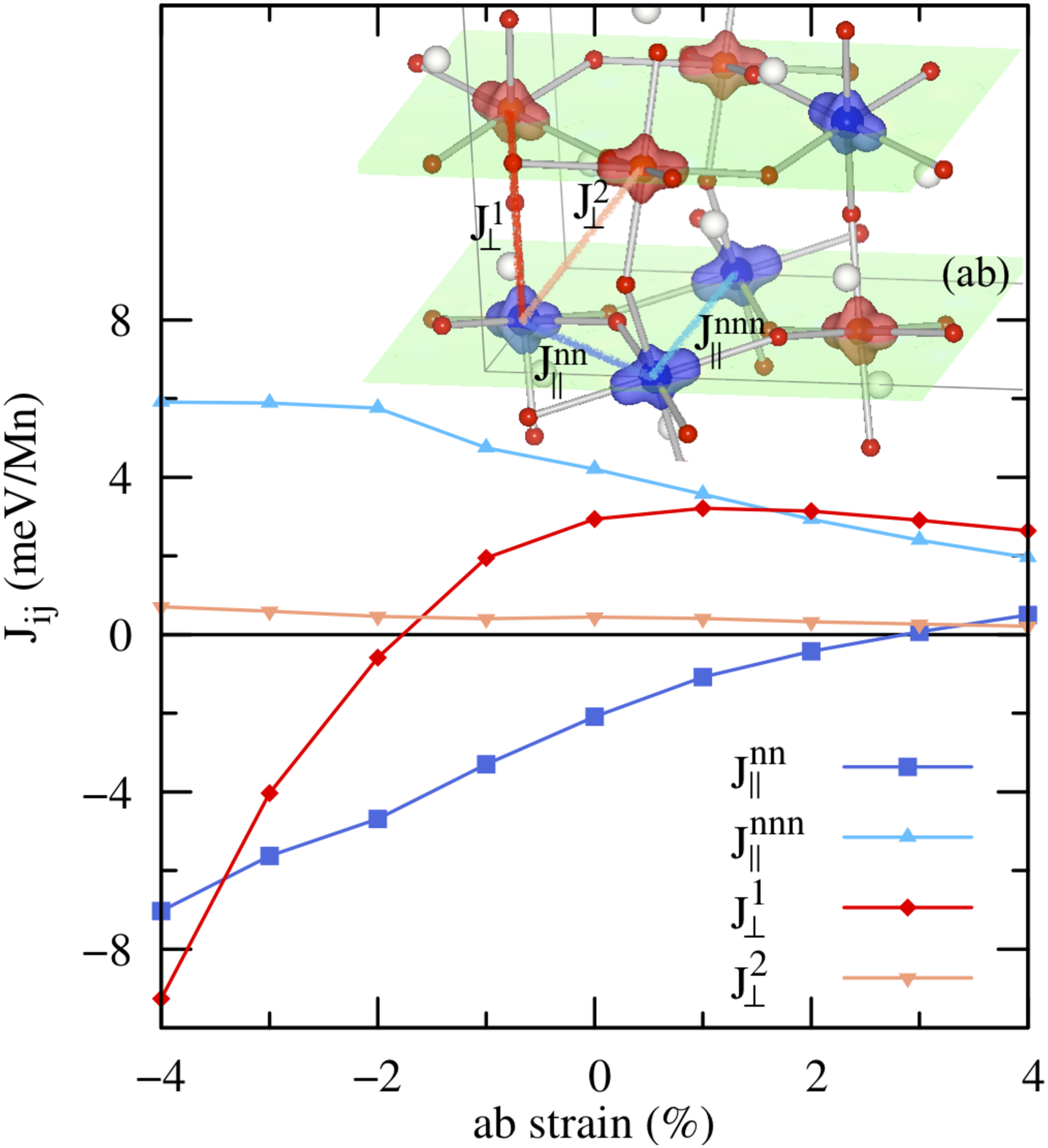}}
\centerline{\includegraphics[width=0.85\columnwidth]{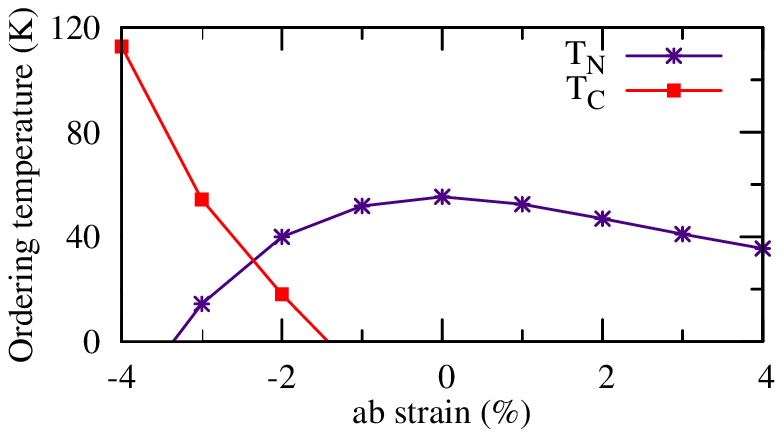}}
\caption{\label{fig.Jij} 
(Color online) Variation of (a) magnetic stabilization (total energy of AFM-A, C, G, E and E$^{*}$ phases with respect to FM in the atomic structure fixed to the optimized AFM-A phase, and the total energy of the AFM-E phase for which the atomic structure was optimized in the non-centrosymmetric phase).
(b) Interatomic exchange interaction ($J_{ij}$):
The first nearest-neighbor in-plane
$J^\textrm{nn}_{\parallel}$,
next nearest-neighbor in-plane
$J^\textrm{nnn}_{\parallel}$,
first nearest-neighbor inter-plane
 $J^{1}_{\bot}$,
next nearest-neighbor inter-plane
 $J^{2}_{\bot}$. 
Positive (negative) sign denotes AFM (FM) coupling.  
(c) Ordering temperatures of the FM and AFM-E phases within the mean-field approximation. 
 }
\end{figure}

In this section we focus on the magnetic stabilization under strain. 
As shown in Fig. \ref{fig.Jij}, the AFM-E phase is stable under tensile strain, but, as a general trend, under compressive strain (at -4\%) none of the AFM phases is any longer stable with respect to the ferromagnetic phase.
This trend is consistent with the evolution of the band gap, which closes towards a metallic state under compressive strain. 
Specifically, the double exchange interaction between the Mn-$e_{g}$ states is enhanced in this region of strain. 
In fact, the in-plane nearest-neighbor FM coupling $J^\textrm{nn}_{\parallel}$ progressively increases under compressive strain. 
The AFM-E configuration, gaining energy both from the double exchange interaction along the zigzag spin chains and from the ferroelectric ionic displacements, is more stable than the FM and AFM-A phases within a region of strain ranging from -3\% to 4\%. 
It is also worth noticing that the inter-plane next-neighbor coupling  $J^{1}_{\bot}$ rapidly changes sign under compressive strain, turning from AFM to FM. 
The reason for the emergence of a FM inter-plane coupling will be explained in the next section. 
From the calculated exchange interactions, $J_{ij}$, we can also estimate the critical temperature $T_c$ in the mean-field approximation for the Heisenberg model: we find that the N\'eel temperature for the AFM-E phase is $k_B\,T_N=2/3(J^\textrm{nn}_{\parallel}+J^{1}_{\bot})$ and that the Curie temperature is $k_B\,T_C=2/3(2J^\textrm{nn}_{\parallel}+J^\textrm{nnn}_{\parallel}+J^{1}_{\bot}+4J^{2}_{\bot})$ for the FM phase, with $k_B$ the Boltzmann constant. Their dependence on strain is shown in Fig. \ref{fig.Jij}.

\section{\label{sec:magn}Ferroelectricity and Jahn-Teller distortion\protect\\}

\begin{figure}[thb] \vspace{3mm}
\centerline{\includegraphics[width=0.85\columnwidth]{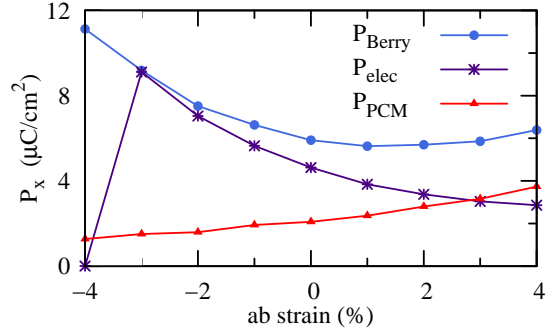}}
\caption{\label{fig.Px} 
(Color online) Variation of the ferroelectric polarization, $P_{\rm Berry}$ (electronic+ionic contribution) and $P_{\rm PCM}$ (ionic contribution), with strain. 
 }
\end{figure}
Finally, we consider the strain effect on the ferroelectric polarization in the AFM-E phase. In order to discuss the evolution of $P$ as a function of strain, we calculate the polarization also at -4\% and -3\% strain, where the AFM-E phase is not magnetically stable with respect to the FM one.
As shown in Fig. \ref{fig.Px}, interestingly,  it turns out that $P$(=P$_{\rm Berry}$) is enhanced at both limits of strain: {\it i.e.} under tensile strain, the contribution to $P$ arising from ionic displacements (=$P_{\rm PCM}$) increases, whereas the electronic contribution to $P$ (=$P_{\rm Berry}$-$P_{\rm PCM}$) decreases, with an opposite trend in the opposite limit of compressive strain. 
The former trend is easily understood: in a larger space, oxygen ions are more freely displaced by the exchange striction, resulting in an enhancement of $P_{\rm ion}$, as discussed in Sec. \ref{sec:struc}. 
The latter is less easy to understand intuitively. 
One possibility is that the hopping of Mn-$e_{g}$ electron increases due to the shorter bond length, so that the {\it asymmetric hopping} is increased, thus explaining the enhanced electronic contribution to the net polarization.
\begin{figure}[tb] \vspace{3mm}
\centerline{\includegraphics[width=0.85\columnwidth]{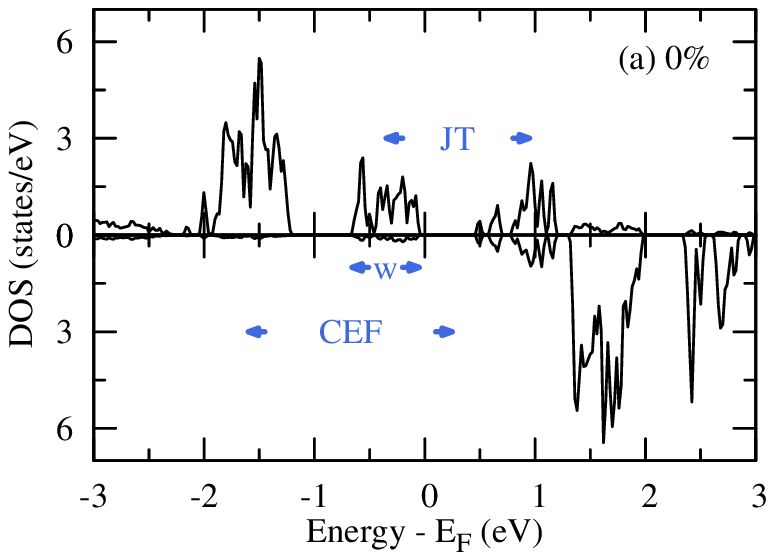}}
\centerline{\includegraphics[width=0.85\columnwidth]{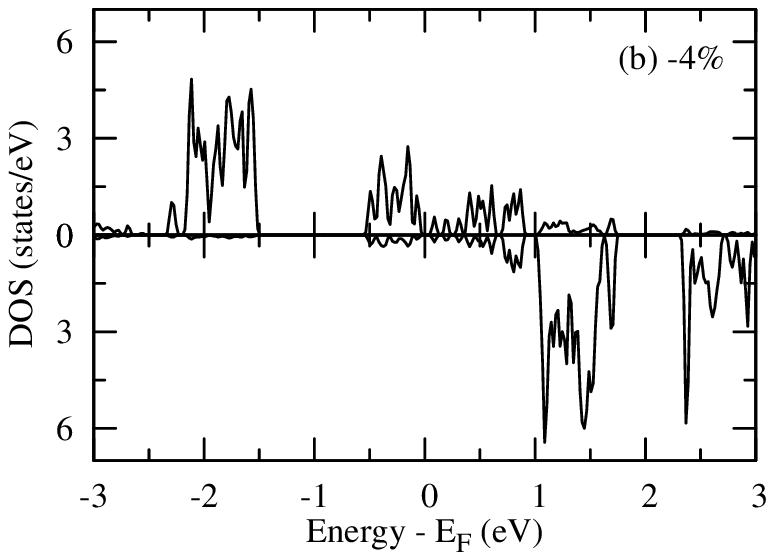}}
\caption{\label{fig.pdos} 
Partial density of states (DOS) at Mn site under no strain and -4\%\ strain. 
JT, w and CEF denote JT splitting at Mn-$e_{g}$ state, band width ($\propto$ hopping integral $t$) of $e_{g}^{1}$ occupied state, and crystal field splitting between $e_{g}$ and $t_{2g}$ states.  
 }
\end{figure}
In Fig. \ref{fig.pdos}, the bandwidth $w$ of Mn-$e_{g}$ states, which reflects the hopping integral between Mn-$e_{g}$ states at neighboring Mn-sites, is shown to be unchanged regardless of strain. 
However, the JT splitting is changed significantly, as one may have expected by looking at the trend of Mn-O bond lengths shown in Fig. \ref{fig.lmsang}, where the long axis $l$ and short axis $s$ get closer under compressive strain. 
Besides, at $-4$\%\ of strain, $l$ and $m$ becomes similar, and as a consequence $Q_{3}$ changes sign from negative to positive, as shown in Fig. \ref{fig.qq}. 
\begin{figure}[htb] \vspace{3mm}
\centerline{\includegraphics[width=0.85\columnwidth]{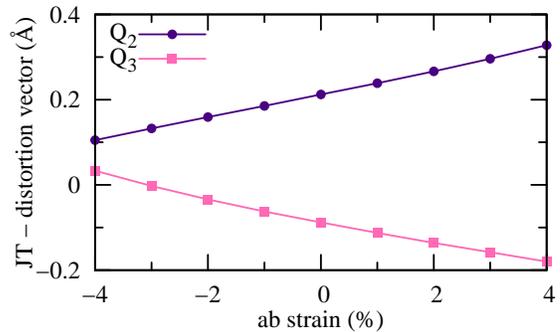}}
\caption{\label{fig.qq} 
(Color online) Variation of the Jahn-Teller vector {\bf Q} = ($Q_{2}$, $Q_{3}$)$\equiv \left( \frac{1}{\sqrt{2}}(l-s), \frac{1}{\sqrt{6}}(2m-l-s) \right)$ \cite{dagotto.review} in the AFM-E configuration, where $l$, $m$ and $s$ are averaged, under in-plane strain.
 }
\end{figure}
This change of the JT distortion is reflected in the shape of the ordered orbitals. 
\begin{figure}[htb] \vspace{3mm}
\centerline{\includegraphics[width=1.00\columnwidth]{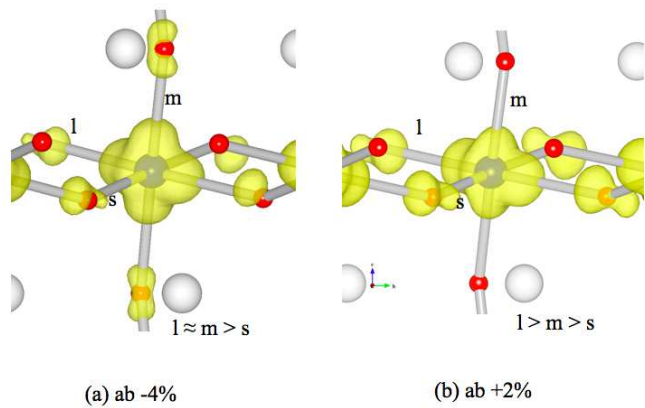}}
\caption{\label{fig.parchg} 
(Color online) Isosurface of charge density of occupied Mn-$e_{g}^{1}$ state  (within an energy range up to 1eV below Fermi energy)
under (a) -4\% and (b) +2\% strain.  
 }
\end{figure}
Figure \ref{fig.parchg} shows the change of the orbital shape under compressive strain. 
The $3x^{2}-r^{2}$ orbital acquires additional character of the $x^{2}-z^{2}$ orbital (in a local frame $x=l$, $y=s$, $z=m$). 
As a consequence of the orbital change, an inter-plane Mn-O bonding through $e_{g}$ electrons is formed (c.f. small weight of the charge density on the apical oxygen atom in Fig.~\ref{fig.parchg}(a) but not in (b)); this might also be related to the fact that inter-plane magnetic coupling turns FM under compressive strain. 
Regarding the enhancement of the electronic contribution to $P$, we should consider the charge transfer caused by Mn electrons at occupied $e_{g}$ state, which hop into unoccupied $e_{g}$ state at neighboring Mn site. 
It is expected that the charge transfer is enhanced when the JT-splitting is small. 
%

\section{Model Hamiltonian }
The change in orbital ordering as a response to compressive strain found in the DFT framework  can be accounted for in a model Hamiltonian approach. 
We consider a degenerate double exchange model, where infinite Hund coupling between the $t_{2g}$ and the $e_g$ electrons is assumed.\cite{dagotto.review} 
This allows us to neglect spin degrees of freedom of conduction electrons, which are fixed by the underlying magnetic structure of the core spins. 
In the E-type configuration, electrons can move along zig-zag chains on a background of parallel (ferromagnetically coupled) core spins, which experience an antiferromagnetic interchain coupling. Orbital ordering can then be stabilized by the
Jahn-Teller coupling to the lattice.  The Hamiltonian reads as:
\begin{eqnarray}\label{ham}
H&=&-\sum_{\langle i,j\rangle} t_{ij}^{\alpha,\beta}\,\cos\Theta_{ij}\,c_{i\alpha}^\dagger\,c_{j,\beta}^{\phantom{\dagger}} +
J \sum_{\langle i,j\rangle}\,\bf{S_{\it i}} \cdot \bf{S_{\it j}} \nonumber \\
&& + E_{JT}\sum_i \left[2(q_{2i}\tau_{xi}+q_{3i}\tau_{zi})+q_{2i}^2+q_{3i}^2 \right].
\end{eqnarray}
We adopt the standard notation, with $c_{i\alpha}^\dagger\,(c_{i\alpha}^{\phantom{\dagger}})$ creation (annihilation) operators for electrons in the bands $\alpha=a,b$ (stemming respectively from $d_{x^2-y^2}\,,\,d_{3z^2-r^2}$ orbitals on Mn$^{3+}$); $t_{ij}^{\alpha,\beta}$ are the oxygen-mediated hopping amplitudes between nearest-neighbors manganese atoms, i.e. $t_{aa}^x=t_{aa}^y=3t_0/4,\,t_{bb}^x=t_{bb}^y=t_0/4,\,t_{ab}^x=-t_{ba}^y=\sqrt{3}t_0/4$ with $t_0=(pd\sigma)^2$ as the energy reference.\cite{slater} 
The $\Theta_{ij}$ are the angles between neighboring Mn-core spins $S_i$ and $S_j$ and $J$ is the exchange coupling which they experience. 
The JT modes are expressed through the dimensionless $q_{2i},q_{3i}\, (=(k/g)\, \mathbf{Q}_{i})$ and $E_{JT}=g^2/2k$ is the static Jahn-Teller energy; $\tau_{\mu i}=\sum_{\alpha\beta}c_{i\alpha}^\dagger\,\sigma_{\alpha\beta}^{\mu}\,c_{i,\beta}^{\phantom{\dagger}}$ are the orbital pseudospins, where $\sigma_{\alpha\beta}^{\mu}$ are the Pauli matrices.

We evaluated the optimal $q_{2i},q_{3i}$ keeping the $t_{2g}$ spins in E-type configuration, and we found that in the undoped case (one electron per Mn site) an orbital ordering occurs as soon as the JT coupling is large enough to induce a finite distortion.
The character of orbitals involved in the ordered phase can be visualized by evaluating the average occupancy of the atomic orbitals, defined as $\rho_{\alpha i}=\langle \phi_{\alpha i}^\dagger \phi_{\alpha i}^{\phantom{\dagger}}\rangle$, where $\phi_{\alpha i}=-\sin(\theta_i/2)\,c_{ai}+\cos(\theta_i/2)\,c_{bi}$ and $\alpha$ labels each of the possible $e_g$-electron orbitals with the corresponding angle $\theta_i$ (e.g. $\theta_i=2\pi/3\,(4\pi/3)$ corresponds to $3x^2-r^2\,(3y^2-r^2)$ orbital, see for instance Ref. \onlinecite{dagotto.review}).

\begin{figure}
\includegraphics[width=1.00\columnwidth]{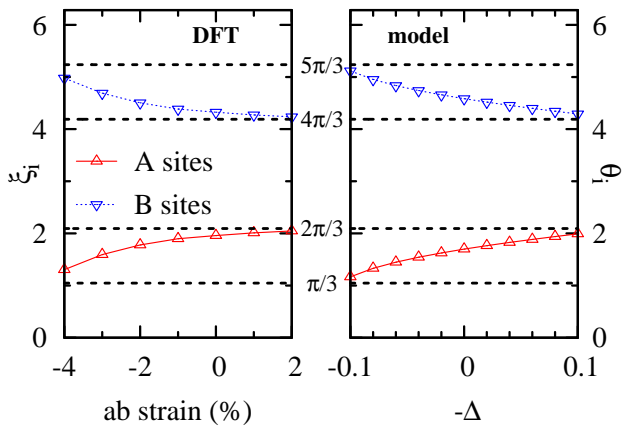}
\caption{(Color online) Jahn-Teller phase $\xi_i$ at neighboring sites as a function of strain as obtained in DFT (left), compared with orbital phase $\theta_i$ evaluated in model Hamiltonian approach as a function of crystal field with reversed sign (right). 
Dotted lines at $\pi/3$ ($2\pi/3$) and $5\pi/3$ ($4\pi/3$) correspond to $x^2-z^2$ ($3x^2-r^2$) and $y^2-z^2$ ($3y^2-r^2$) orbitals respectively. }\label{fig2}
\end{figure}

The effect of the in-plane strain can be mimicked by the introduction of a crystal-field term $+\Delta\,\sum_i\tau_{zi}$ in the Hamiltonian (\ref{ham}); if we consider a compression in the $ab$ plane, then $\Delta$ should be chosen positive. 
Another effect of compression is of course to increase the elastic stiffness of MnO$_6$ octahedra, therefore one should expect a decrease of JT static energy (as indeed observed in the reduction of the JT gap found in DFT calculations). 
We found that, upon increasing $\Delta$, the orbital ordering is not affected in its alternating pattern, but there is a change in the weight of involved atomic orbitals, namely the weight of $x^2-z^2$ ($y^2-z^2$) is increased with respect to $3x^2-r^2$ ($3y^2-r^2$). 
At the same time, the gap between the $e_g$-electron bands is reduced even with fixed JT coupling. 
To visualize the evolution of orbitals involved in the ordered pattern as a function of strain we can evaluate the angle $\theta_i=\tan^{-1}(\langle\tau_{xi}\rangle/\langle\tau_{zi}\rangle)$ and compare it with the Jahn-Teller phase
$\xi_i=\tan^{-1}(Q_{2i}/Q_{3i})$ extrapolated from DFT calculations. 
As shown in Fig. \ref{fig2}, we find a good qualitative agreement between the two phases.

We have also checked our results by considering the effect of cooperative distortions of oxygens. 
This has been done by optimizing the Hamiltonian (\ref{ham}) with respect to oxygen displacements $\{u\}$ from their equilibrium positions rather than optimizing the local distortions of the MnO$_6$ octahedron labeled by $\{q\}$: in this case the strain can be mimicked by changing the equilibrium positions of oxygens, specifically decreasing the MnO distances in the $ab$ plane with respect to the equilibrium position of oxygens along the $c$ axis.
This approach provides results which are consistent with those obtained in the noncooperative treatment of Jahn-Teller effect.

\begin{figure}
\includegraphics[width=0.85\columnwidth]{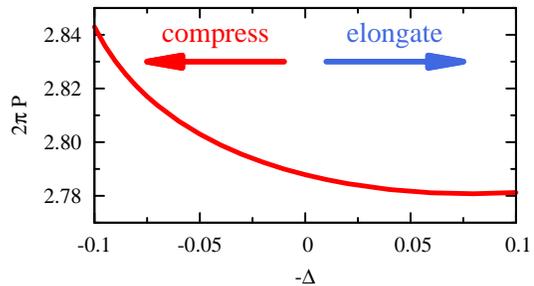}
\caption{(Color online) Electronic polarization as a function of -$\Delta$ ($\Delta$ is the crystal field).}\label{fig3}
\end{figure}

We focus then on possible electronic polarization. Due to the AFM-E magnetic configuration, at $T=0$ $e_g$ electrons can only hop between sites with ferromagnetically aligned core-spins. 
This implies that hopping processes can occur only along one-dimensional zig-zag chains. 
As discussed in Ref. \onlinecite{koizumi}, the presence of Jahn-Teller centers in this one-dimensional system can induce a Berry phase in the Bloch function, intimately related to the orbital ordering induced by JT distortions, from which electronic polarization can be inferred. We follow the prescription described in Ref. \onlinecite{resta} and evaluate the polarization within the chains as
$\displaystyle{P_{el}=\lim_{L\to\infty}\,\frac{e}{2\pi}\,\mbox{Im}\ln\langle\Psi_0\vert\,e^{i\frac{2\pi}{L}\hat{X}}\vert\Psi_0\rangle,
}$
where $L$ is the length of the one-dimensional chain (thermodynamic limit is taken at the end of the calculation),$\Psi_0$ is the ground-state wavefunction of model (\ref{ham}) with optimized $q_{2i},q_{3i}$, and $\hat{X}=\sum_i\,i\,\hat{n}$ is the electronic position operator. 
The onset of the orbital ordering within the zigzag chains induces a finite polarization whose origin is fully electronic, the ions position being fixed in the system.\cite{PB}
As shown in Fig. \ref{fig3}, we find that $P_{el}$ is indeed enhanced by increasing $\Delta$ (i.e. as a consequence of a compressive strain in our model approach), in excellent qualitative agreement with the DFT results.

\section{Conclusions}
The ferroelectric and magnetic properties of orthorhombic \hmo~were investigated by a combined density functional theory and model Hamiltonian study. 
We found an increase in the electric polarization when the system is subjected to in-plane compressive strain. 
This is closely related to the modifications in the electronic structure, namely the orbital ordered states $3x^2-r^2/3y^2-r^2$ gain additional character of the $x^2-z^2/y^2-z^2$ states. 
Beside leading to an increase of the polarization, the orbital mixing influences the magnetic stability; thus, a crossover from the AFM-E to the FM phase occurs around -3\% compressive strain.
\acknowledgments
The research leading to these results has received funding from the European Research Council under the EU Seventh Framework Programme  (FP7/2007-2013) / ERC grant agreement n. 203523. 
Computational support from the Swedish National Infrastructure for ScientiÞc Computing (SNIC), Caspur  Supercomputing Center  (Rome) and Cineca Supercomputing Center (Bologna) is gratefully acknowledged. 
B.~S. acknowledges support from VR. O.~E. acknowledges support from VR, SSF, ERC, and the European Comission, project number FP7-NMP-2008-EU-India2-233513.
The charge density plots were performed using the program VESTA.\cite{vesta}

\end{document}